\documentclass[twocolumn,secnumarabic,amssymb, amsmath, nofootinbib,tightenlines, nobibnotes, aps, prl,epsfig]{revtex4}
\usepackage{graphicx}% Include figure files
\usepackage{dcolumn}% Align table columns on decimal point
\usepackage{bm}% bold math
\begin{document}
\preprint{APS/123-QED}
\title{ Nonlinear Corrections to the Momentum Sum Rule}% Force line breaks with \\

\author{G.R.Boroun}%
 \email{boroun@razi.ac.ir }
%\author{B.Rezaei }
%\altaffiliation{brezaei@razi.ac.ir}%Lines break automatically or can be forced with \\

\affiliation{Department of  Physics, Razi University, Kermanshah
67149, Iran}% \textbackslash\textbackslash

%\author{D.Schildknecht}%
%\email{schild@physik.uni-bielefeld.de } \affiliation{
%Fakult$\ddot{a}$t f$\ddot{u}$r Physik, Universit$\ddot{a}$t
%Bielefeld,
%D-33501 Bielefeld, Germany}% \textbackslash\textbackslash
\date{\today}% It is always \today, today,
\begin{abstract}
%%%%%%%%%%%%%%%%%%%%%%%%%%%%%%%%%%%%%%%%%%%%%%%%%%%%%%%
The importance of the nonlinear corrections on the momentum sum
rule is investigated on the initial scale $Q_{0}^2$. Nonlinear
corrections are found to play an indispensable role in the singlet
and gluon momentum sum rule in the high-order approximations in
the parameterization groups for nucleons and light nuclei at low
$x$ in future colliders. In this way, we obtain a significantly
different low $x$ behavior of the singlet and gluon momentum sum rule at the hotspot point.\\
%%%%%%%%%%%%%%%%%%%%%%%%%%%%%%%%%%%%%%%%%%%%%%%%%%%%%%%
\end{abstract}
 \pacs{***}%PACS, the Physics and Astronomy
                              %Classification Scheme.
 %Use showkeys class option if keyword
\maketitle
%**********************************************************
%%%%%%%%%%%%%%%%%%%%%%%%%%%%%%%%%%%%%%%%%%%%%%%%%%%%%%%%%%%%%%%%%%%%%%%%%%%%%%%%%%%%%%%%%%%%%
\subsection{1. Introduction}

Sum rules are integrals over structure functions or parton
distributions, which are extremely useful in calculating the
lowest-mass hadronic bound states or determining effective
coupling constants. The starting point for QCD sum rules is the
operator-product expansion (OPE), which was formulated firstly by
M.A.Shifmann, A.I.Vainshtein and V.I.Zakharov in Ref.[1]. The QCD
sum rules are a phenomenological procedure for evaluating the
matrix elements of the operators that occur, and OPE gives a
general form for the quantities of interest [2]. The matrix
elements of the operators in the OPE for the forward virtual
Compton amplitude $\gamma^{*}p{\rightarrow}\gamma^{*}p$ correspond
with the Mellin moments of the structure functions [3]. Sometimes
the origin of a sum rule is more fundamental than the quantum
parton model (QPM). We might call a sum rule an exact QCD sum rule
if its result found within the context of the parton model is not
altered by any radiative or non-perturbative correction [4,5]. The
first one is the Adler [6] sum rule, where it is for the charged
current structure functions and follows from the current
conservation. The Bjorken [7] and Gross-Llewellyn [8] sum rules
get a QCD correction, where the parton model results are modified
by radiative corrections. The Gottfried [9] sum rule requires the
assumption of an SU(2) symmetric sea and is affected by
non-perturbative physics.\\
The momentum sum rule (MSR) has been seen as a convenient
constraint on the definition of the parton distribution functions
(PDFs) rather than a basic QCD sum rule when going beyond leading
order QCD. The sum of the fraction of the proton$^,$s momentum
carried by quarks must be less than unity and the remaining
momentum is carried by the gluons. Botje in Ref.[10] was shown
that integral over the PDFs for quarks and gluons at the initial
scale $Q_{0}^2$ gives $
\int_{0}^{1}dx{x}\sum_{q}(q(x)+\overline{q}(x))=0.594{\pm}0.018$
and $\int_{0}^{1}dx{x}g(x)=0.394{\pm}0.018$ where
$\sum_{q}(q+\overline{q})$ is the flavor-singlet contribution and
$g$ is the gluon distribution. In fact the gluons carry the very
large missing fraction of the proton momentum. The MSR from the
second Mellin moment reads
\begin{eqnarray}
\int_{0}^{1}dx{x}\bigg{[}\sum_{q}(q(x,Q_{0}^{2})+\overline{q}(x,Q_{0}^{2}))+g(x,Q_{0}^{2})\bigg{]}=1,
\end{eqnarray}
and modified by the following form
\begin{eqnarray}
\int_{0}^{1}dx{x}\bigg{[}u_{v}(x,Q_{0}^{2})+d_{v}(x,Q_{0}^{2})+S(x,Q_{0}^{2})+g(x,Q_{0}^{2})\bigg{]}=1,
\end{eqnarray}
where the light quark sea contribution is defined as
$S{\equiv}2(\overline{u}+\overline{d})+s+\overline{s}$.\\
In order to make a precise determination of parton distribution
sets (such as CTEQ, nCTEQ, MSTW, GRV, GJR and NNPDF
Collaborations), one has to use a large set of data which together
cover a large range of $x$ and $Q^2$ and put stringent constraints
on the various parton types within the proton. Usually the
following parameterizations for gluon and sea quark distributions
are used by the above groups at the initial scale $Q_{0}^{2}$. As
an example we have
\begin{eqnarray}
xf_{i}(x,Q_{0}^{2})=A_{i}x^{\delta_{i}}(1-x)^{\eta_{i}}(1+\epsilon_{i}\sqrt{x}+\gamma_{i}x).
\end{eqnarray}
The values of the parameters obtained from a global QCD fit and
also the normalization parameters are fixed by the MSR and valence
quark counting
rules\footnote{$\int_{0}^{1}dx~u_{v}(x,Q_{0}^{2})=2$ and
$\int_{0}^{1}dx~d_{v}(x,Q_{0}^{2})=1$.}. Differences in the
results of different groups are made from many subjective choices
in the selection of data sets and kinematic boundaries, together
with the choice of the strong coupling constant and the correction
for non-perturbative and target mass and  nuclear effects of data
from fixed target experiments [4].\\
In this paper we consider the nonlinear corrections (NLCs) to the
momentum sum rule and show differences in the MSR due to PDF sets
in LO, NLO and NNLO at $\alpha_{s}$ for nucleons and nuclei. In
the next section, the theoretical formalism is presented,
including the nonlinear corrections at the initial scale.\\

\subsection{2. Nonlinear Corrections}

The nonlinear corrections (or gluon recombination effects) are not
negligible in the low $x$, low $Q^{2}$ region and it is known that
this behavior reduces the growth of the gluon distribution. These
nonlinear effects were defined by Gribov-Levin-Ryskin [11] and
Mueller-Qiu [12] (GLR-MQ). The main difference of this equation
from the Linear evolution equation is the presence of the
quantity, $G^2$ which is interpreted as the two-gluon distribution
per unit area of hadron. The dominant source for the nonlinear
corrections at low $x$ is the conversion of the two gluon ladders
merged into a gluon or a quark-antiquark pair, as the evolution of
the parton distributions is directly related to the gluon-gluon
fusion terms in the GLR-MQ evolution equations [13,14]. Indeed,
this leads to saturation of the gluon density at low $Q^{2}$ with
decreasing $x$, when $W{\lesssim}\alpha_{s}$ where
$W=\frac{n_{g}\widehat{\sigma}}{\pi{R^2}}{\sim}\frac{\alpha_{s}(Q^2)}{\pi{R^2}
Q^2}xg(x,Q^2)$. Here, $n_{g}$ is the number of gluons,
$\widehat{\sigma}$ is the gluon-gluon cross section and $\pi{R^2}$
is the transverse area of a hadron where $R$ is the characteristic
radius of the gluon distribution in the hadronic target, which
determines the strength of the nonlinear corrections and it comes
from the integration over the transverse components of $k$ ($k$ is
the transverse momenta of gluons),
$\frac{1}{R^2}{\sim}\int{dk^{2}_{T}}[F(-k^{2}_{T})]^2$. When gluon
ladders are coupled to the proton radius, then the value of $R$ is
given by $R=5~\mathrm{GeV}^{-1}$. Indeed, the form factor $F$ is
characterized by the proton radius. The value
$R=2~\mathrm{GeV}^{-1}$ signifies the gluons concentrated on the
hotspots [15-17]. Although a more precise nonlinear evolution
equation was developed by Balitsky-Kovchegov [18,19] based on the
evolution of BFKL, but to study the possible importance of
shadowing we base our starting gluon and singlet distributions
$g(x,Q_{0}^{2})$ and $S(x,Q_{0}^{2})$ on the solutions of the
GLR-MQ evolution equation based on Refs.[15,16].\\
On the basis of the shadowing effects, the nonlinear corrections
to the MSR read
\begin{eqnarray}
\int_{0}^{1}dx{x}\bigg{[}u_{v}(x,Q_{0}^{2})+d_{v}(x,Q_{0}^{2})~~~~~~~~~~~~~~~~~~~~~~~~~\nonumber\\
~~~~~~~~~~~~~~+S^{\mathrm{NLC}}(x,Q_{0}^{2})+g^{\mathrm{NLC}}(x,Q_{0}^{2})\bigg{]}=1,
\end{eqnarray}
where
\begin{eqnarray}
xg^{\mathrm{NLC}}(x,Q_{0}^2)=xg(x,Q_{0}^2)\xi^{NLC}(x,x_{0},Q_{0}^2),
\end{eqnarray}
and
\begin{eqnarray}
\xi^{\mathrm{NLC}}(x,x_{0},Q_{0}^2)&=&\Big{\{}1+\theta(x_{0}-x)
\Big{[}
xg(x,Q_{0}^2)\nonumber\\
&&-xg(x_{0},Q_{0}^2)
\Big{]}/xg_{\mathrm{sat}}(x,Q_{0}^{2})\Big{\}}^{-1},
\end{eqnarray}
with
\begin{eqnarray}
xg_{\mathrm{sat}}(x,Q^{2})=\frac{16{R}^{2}Q^2}{27{\pi}\alpha_{s}(Q^2)},
\end{eqnarray}
where $g_{\mathrm{sat}}$ is the value of the gluon which would
saturate the unitarity limit in the leading shadowing
approximation. The parameter $x_{0}$ is introduced to be
$x_{0}{\simeq}10^{-2}$ so that the nonlinear corrections are
negligible for $x{\geq}x_{0}$. The nonlinear corrections to the
gluon distribution are reflected in the sea-quark distributions
where the sea-quark starting distribution in the region $x<x_{0}$
is proportioned to the nonlinear correction to the gluon by the
following form [15,16]
\begin{eqnarray}
xS^{\mathrm{NLC}}(x,Q_{0}^2)=xS(x,Q_{0}^2)\xi^{NLC}(x,x_{0},Q_{0}^2).
\end{eqnarray}
Therefore the NLCs to the MSR is defined by
\begin{eqnarray}
\int_{0}^{1}dx{x}\bigg{[}u_{v}(x,Q_{0}^{2})+d_{v}(x,Q_{0}^{2})~~~~~~~~~~~~~~~~~~~~~~~~\nonumber\\
~~~+\xi^{\mathrm{NLC}}(x,x_{0},Q_{0}^2)\bigg{\{}S(x,Q_{0}^{2})+g(x,Q_{0}^{2})\bigg{\}}\bigg{]}=1.
\end{eqnarray}
It is useful to consider the linear and nonlinear corrections to
the singlet and gluon distribution functions in which the kernel
$P_{s+g}$ reads
\begin{eqnarray}
P^{\mathrm{NLC}}_{s+g}(x,Q_{0}^{2})&=&\xi^{\mathrm{NLC}}(x,x_{0},Q_{0}^2)(xS(x,Q_{0}^{2})\nonumber\\
&&+xg(x,Q_{0}^{2})),
\end{eqnarray}
where the momentum sum rule for single and gluon distributions is
defined by
\begin{eqnarray}
I^{\mathrm{NLC}}_{s+g}=\int_{0}^{1}P^{\mathrm{NLC}}_{s+g}(x,Q_{0}^{2})dx.
\end{eqnarray}
The NLCs to the MSR in a nucleus are modified owing to by the
nuclear modification factor $w_{i}(x,A,Z)$, as
\begin{eqnarray}
\int_{0}^{1}dx{x}\bigg{[}u^{A}_{v}(x,Q_{0}^{2})+d^{A}_{v}(x,Q_{0}^{2})~~~~~~~~~~~~~~~~~~\nonumber\\
~+\xi_{A}^{\mathrm{NLC}}(x,x_{0},Q_{0}^2)\bigg{\{}S^{A}(x,Q_{0}^{2})+g^{A}(x,Q_{0}^{2})\bigg{\}}\bigg{]}=1,
\end{eqnarray}
where the initial nuclear parton distributions at a fixed
$Q_{0}^{2}$ are defined by the following form
\begin{eqnarray}
f^{A}_{i}(x,Q_{0}^{2})=w_{i}(x,A,Z)f_{i}(x,Q_{0}^{2}),
\end{eqnarray}
where the modification function, based on the QCD analysis, in a
cubic type is
\begin{eqnarray}
w_{i}(x,A,Z)=1+\Big{(}1-\frac{1}{A^{\alpha}}\Big{)}\frac{a_{i}(A,Z)+H_{i}(x)}{(1-x)^{\beta_{i}}},
\end{eqnarray}
where $H_{i}(x)=b_{i}(A)x+c_{i}(A)x^2 +d_{i}(A)x^3$  is available
in the literature [20-23]. The value of $R^{A}$ for a nuclear
target with the mass number $A$, in the nuclear gluon saturation
$xg^{A}_{\mathrm{sat}}$, is defined by $R^{A}=A^{1/3}R$ [24]. The
momentum sum rule for single and gluon distributions in nuclei
modified by the following form
\begin{eqnarray}
I^{\mathrm{NLC}}_{sA+gA}=\int_{0}^{1}P^{\mathrm{NLC}}_{sA+gA}(x,Q_{0}^{2})dx
\end{eqnarray}
where
\begin{eqnarray}
P^{\mathrm{NLC}}_{sA+gA}(x,Q_{0}^{2})&=&\xi_{A}^{\mathrm{NLC}}(x,x_{0},Q_{0}^2)(xS^{A}(x,Q_{0}^{2})\nonumber\\
&&+xg^{A}(x,Q_{0}^{2})).
\end{eqnarray}
The results that are obtained in the above (i.e., Eqs.11 and 16)
can be confirmed by simulating the events in $eA$ collisions at
the large hadron electron collider (LHeC) [25] and the
Electron-Ion Collider (EIC) [26] energies in future.\\

\subsection{3. Results and Conclusions}

We begin by specifying the parametrization of the singlet and
gluon distributions via the global parton analysis\footnote{The
parton distributions are determined in the NLO and NNLO
approximations from the global analysis of hard-scattering data
within the standard framework of leading-twist fixed-order
collinear factorization in the $\overline{\mathrm{MS}}$ scheme
[27].} of MSTW 2008 collaboration, where the optimal values of
$\alpha_{s}$ and the input singlet and gluon parameters at
$Q_{0}^{2}=1~\mathrm{GeV}^2$ in the LO up to NNLO approximations
are determined and summarized in Ref.[27]. The values
$\alpha_{s}(M_{Z}^{2})$ at the LO, NLO and NNLO approximations are
the values of 0.13939, 0.12018 and 0.11707 respectively. The new
singlet and gluon distributions from the CT18
collaboration\footnote{The new PDFs from the CTEQ-TEA
collaboration, obtained using a wide variety of high-precision
Large Hadron Collider (LHC) data, in addition to the combined HERA
I+II deep-inelastic scattering data set in the NNLO approximation
[28].} are used at the initial scale
$Q_{0}^{2}=1.69~\mathrm{GeV}^2$ in the NNLO approximation where in
Ref.[28] $\alpha_{s}(M_{Z}^{2})=0.1164$. Nuclear singlet and gluon
distribution functions\footnote{The nuclear deep-inelastic
scattering data analyzed in Ref.[29] are complemented by the
available charged current neutrino DIS data with nuclear targets
and data from Drell-Yan cross section measurements for several
nuclear targets in the NLO and NNLO approximations.} in a general
mass variable flavor number scheme have been called from Ref.[29],
where the CT18 PDFs are used as baseline proton PDFs and the
strong coupling constant is taken as
$\alpha_{s}(M_{Z}^{2})=0.118$. In Ref.[30], the singlet and gluon
distributions from JR09 collaboration\footnote{The deep inelastic
scattering and Drell-Yan dimuon production data in the NNLO
approximation are available [30].} in the NNLO approximation at
the initial scale $Q_{0}^{2}=2~\mathrm{GeV}^2$ with
$\alpha_{s}(M_{Z}^{2})=0.1158$. The nPDFs\footnote{The nuclear
parton distribution functions (nPDFs) are obtained from neutral
current charged-lepton deeply inelastic scattering  data and
Drell-Yan (DY) cross-section ratios
$\sigma^{A}_{DY}/\sigma^{A'}_{DY}$ for several nuclear targets in
the NNLO approximation [21]. } at the NNLO approximation are
obtained from Refs.[21,22] where all parton distributions are
obtained from JR09 [30] and CT18 [28] set of the free proton PDFs,
respectively.\\
Figures 1 and 2 show the kernel $P_{s+g}(x,Q_{0}^{2})$ (i.e.,
Eq.(10)) for the linear
($\xi^{{\mathrm{NLC}}}(x,x_{0},Q_{0}^{2})=1$) correction at the
input scale $Q_{0}^{2}=2~\mathrm{GeV}^2$ and $1.69~\mathrm{GeV}^2$
in a wide range of $x$ for the JR09 [30] and CT18 [28]
parameterization models  in the NNLO approximation, respectively.
The nonlinear ($\xi^{{\mathrm{NLC}}}(x,x_{0},Q_{0}^{2}){\neq}1$)
corrections are compared with the linear in the region
$x{\leq}10^{-2}$ for both $R=2~\mathrm{GeV}^{-1}$(dashed-dot
curves) and $R=5~\mathrm{GeV}^{-1}$(dashed curves) in Figs.1 and
2. We observe that the kernel $P_{s+g}(x,Q_{0}^{2})$ is violated
[31] at low $x$ if we consider the nonlinear corrections at the
hotspot. These violations are visible at low $x$ for the JR09 and
CT18  parameterization models. As we see from Figs.1 and 2,
accounting for nonlinear corrections gives a noticeably larger
linear singlet+gluon at low $x$ at the hotspot point.\\
\begin{figure}
\centerline{
\includegraphics[width=0.5\textwidth]{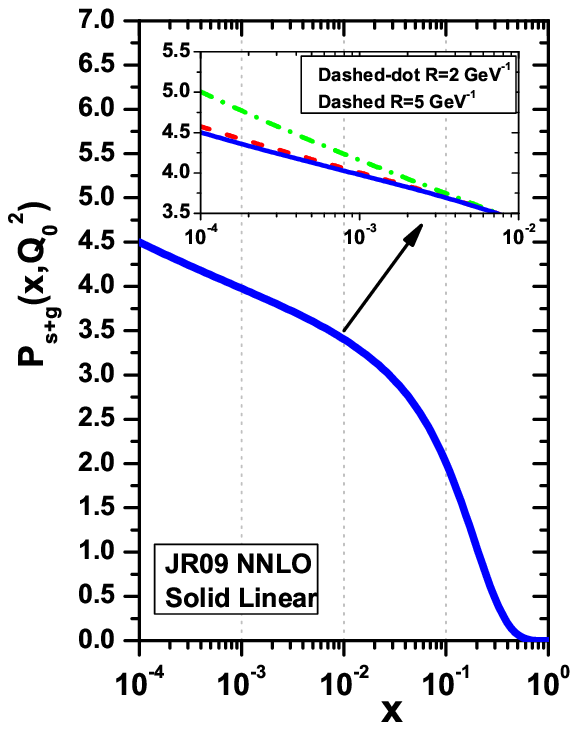}}
\caption{ $P_{s+g}(x,Q_{0}^{2})$ as a function of $x$ on the
initial scale $Q_{0}^{2}=2~\mathrm{GeV}^2$ for the JR09
parameterization model [30] in the NNLO approximation. The inset:
the nonlinear corrections compared with the linear (solid curve)
at $R=5~\mathrm{GeV}^{-1}$ (dashed curve) and
$R=2~\mathrm{GeV}^{-1}$ (dashed-dot curve) for
$x{\leq}10^{-2}$.}\label{Fig1}
\end{figure}
\begin{figure}
\centerline{
\includegraphics[width=0.5\textwidth]{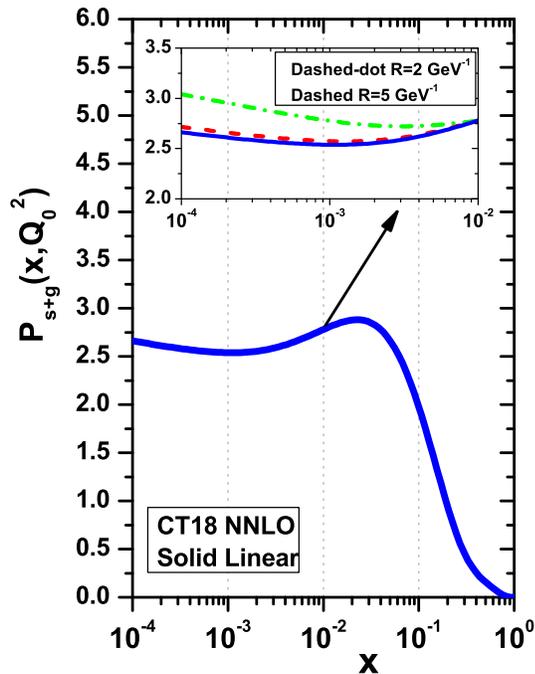}}
\caption{ $P_{s+g}(x,Q_{0}^{2})$ as a function of $x$ on the
initial scale $Q_{0}^{2}=1.69~\mathrm{GeV}^2$ for the CT18
parameterization model [28] in the NNLO approximation. The inset:
the nonlinear corrections compared with the linear (solid curve)
at $R=5~\mathrm{GeV}^{-1}$ (dashed curve) and
$R=2~\mathrm{GeV}^{-1}$ (dashed-dot curve) for
$x{\leq}10^{-2}$.}\label{Fig2}
\end{figure}
In Tables I and II, the singlet and gluon contributions to the MSR
(i.e., $I_{s+g}$) by the nonlinear corrections at
$R=5~\mathrm{GeV}^{-1}$ and $R=2~\mathrm{GeV}^{-1}$ are compared.
We compare the linear of $I_{s+g}$ with the nonlinear corrections
for the MSTW parameterization model [27] in the LO, NLO and NNLO
approximations on the initial scale $Q_{0}^{2}=1~\mathrm{GeV}^2$.
In Table II, we compared the linear and nonlinear corrections to
the  singlet and gluon contributions of the MSR on the initial
scale $Q_{0}^{2}=2~\mathrm{GeV}^2$ for the JR09 parameterization
model [30] and on the initial scale
$Q_{0}^{2}=1.69~\mathrm{GeV}^2$ for the CT18 parameterization
model [28]  in the NNLO approximation respectively, at
$R=5~\mathrm{GeV}^{-1}$ and $R=2~\mathrm{GeV}^{-1}$. The
differences between the linear and nonlinear corrections for the
singlet and gluon contributions in the momentum sum rule are
defined by the following form
\begin{eqnarray}
\Delta=I^{\mathrm{Linear}}_{s+g}-I^{\mathrm{Nonlinear}}_{s+g}(R).
\end{eqnarray}
We observe in Tables I and II that the nonlinear corrections to
the MSR increase the singlet and gluon contributions\footnote{For
all parameterization groups considered.} in comparison with the
linear at the initial scales. Indeed, the momentum carried by the
singlet and gluon distributions increases when we consider the
nonlinear corrections to the distribution functions at the initial
scales.
\begin{table}[h]
\centering \caption{$I_{s+g}$ on the initial scale
$Q_{0}^{2}=1~\mathrm{GeV}^2$ for the MSTW parameterization model
[27] in the LO, NLO and NNLO approximations. The nonlinear
corrections obtained at $R=5~\mathrm{GeV}^{-1}$  and
$R=2~\mathrm{GeV}^{-1}$ respectively. The differences between the
linear and nonlinear corrections for the singlet and gluon
contributions in the momentum sum rule are determined.
}\label{table:table1}
\begin{minipage}{\linewidth}
\renewcommand{\thefootnote}{\thempfootnote}
\centering
\begin{tabular}{|l||c|c|c|} \hline\noalign{\smallskip} $I_{s+g}$ & LO & NLO & NNLO   \\
\hline\noalign{\smallskip}
Linear                              & 0.5219817 & 0.5150395 & 0.5137795  \\
Nonlinear($R=5~\mathrm{GeV}^{-1}$)          & 0.5213362 & 0.5151779 & 0.5138630  \\
Nonlinear($R=2~\mathrm{GeV}^{-1}$)          & 0.5222538 & 0.5193943 & 0.5095848   \\
$\Delta$($R=5~\mathrm{GeV}^{-1}$)          & 0.0006455 & -0.0001385 & -0.000834   \\
$\Delta$($R=2~\mathrm{GeV}^{-1}$)          & -0.0002721 & -0.0043549 & 0.0041947   \\
\hline\noalign{\smallskip}
\end{tabular}
\end{minipage}
\end{table}
\begin{table}[h]
\centering \caption{$I_{s+g}$ on the initial scale
$Q_{0}^{2}=2~\mathrm{GeV}^2$ for the JR09 parameterization model
[30] and on the initial scale $Q_{0}^{2}=1.69~\mathrm{GeV}^2$ for
the CT18 parameterization model [28]  in the NNLO approximation
respectively. The nonlinear corrections obtained at
$R=5~\mathrm{GeV}^{-1}$ and $R=2~\mathrm{GeV}^{-1}$ respectively.
The differences between the linear and nonlinear corrections for
the singlet and gluon contributions in the momentum sum rule are
determined. }\label{table:table1}
\begin{minipage}{\linewidth}
\renewcommand{\thefootnote}{\thempfootnote}
\centering
\begin{tabular}{|l||c|c|} \hline\noalign{\smallskip} $I_{s+g}$ & JR09 & CT18    \\
\hline\noalign{\smallskip}
Linear                              & 0.5451307 & 0.5404662   \\
Nonlinear($R=5~\mathrm{GeV}^{-1}$)          & 0.5452354 & 0.5406253  \\
Nonlinear($R=2~\mathrm{GeV}^{-1}$)          & 0.5458140 & 0.5404662    \\
$\Delta$($R=5~\mathrm{GeV}^{-1}$)          & -0.0001047 & -0.0001591   \\
$\Delta$($R=2~\mathrm{GeV}^{-1}$)          & -0.0006833 & -0.0010523    \\
\hline\noalign{\smallskip}
\end{tabular}
\end{minipage}
\end{table}
By having these corrections at low $x$, it will be possible to
redefine the MSR for the parameters on the singlet and gluon
distributions for all sets of parametrizations at the starting
scale $Q_{0}^{2}$ in future colliders. This provides a mechanism
regulating the respective amounts of nonlinear corrections  at low
$x$ [32].\\
A main observable consequence is extracting nonlinear corrections
to the momentum sum rule for nuclei\footnote{An interesting novel
correction to the MSR for nuclear structure functions is addressed
by the authors in Ref.[32]. }. Figures 3-5 show the kernel
$P_{sA+gA}(x,Q_{0}^{2})$ (i.e., Eq.(16)) for the linear
($\xi_{A}^{{\mathrm{NLC}}}(x,x_{0},Q_{0}^{2})=1$) correction at
the input scale $Q_{0}^{2}=2~\mathrm{GeV}^2$ and
$1.69~\mathrm{GeV}^2$ in a wide range of $x$ for the JR09 [30] and
CT18 [28] parameterization models  in the NNLO approximation,
respectively, where the nuclear modifications are provided by a
weight function\footnote{The nuclear PDFs are related to the PDFs
in a free proton by multiplying  a weight function $w_{i}(x,A,Z)$
at the input scale. } $w_{i}(x,A,Z)$ by the KT16 [21] and KSTSG21
[29] respectively.
\begin{figure}
\centerline{
\includegraphics[width=0.5\textwidth]{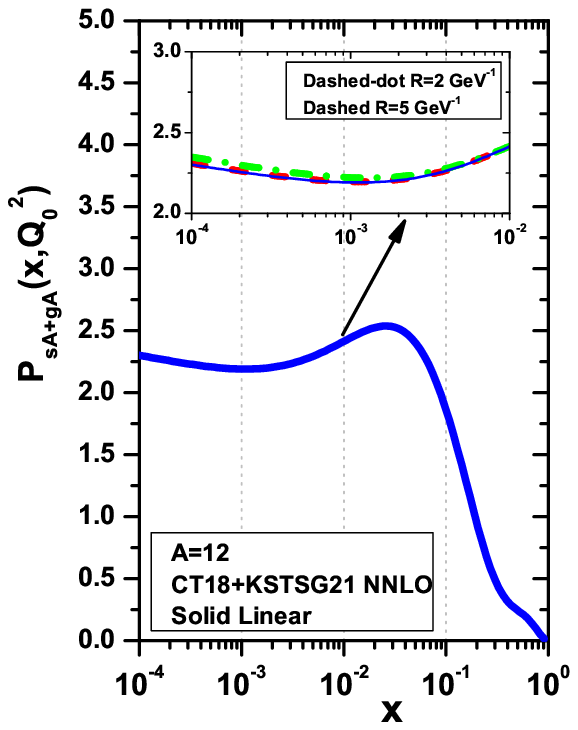}}
\caption{ $P_{sA+gA}(x,Q_{0}^{2})$ as a function of $x$ on the
initial scale $Q_{0}^{2}=1.69~\mathrm{GeV}^2$ for the CT18 [28]
and KSTSG21 [29] parameterization models in the NNLO approximation
of the nucleus of C-12(A=12,Z=6). The inset: the nonlinear
corrections compared with the linear (solid curve) at
$R=5~\mathrm{GeV}^{-1}$ (dashed curve) and $R=2~\mathrm{GeV}^{-1}$
(dashed-dot curve) for $x{\leq}10^{-2}$.}\label{Fig3}
\end{figure}
\begin{figure}
\centerline{
\includegraphics[width=0.5\textwidth]{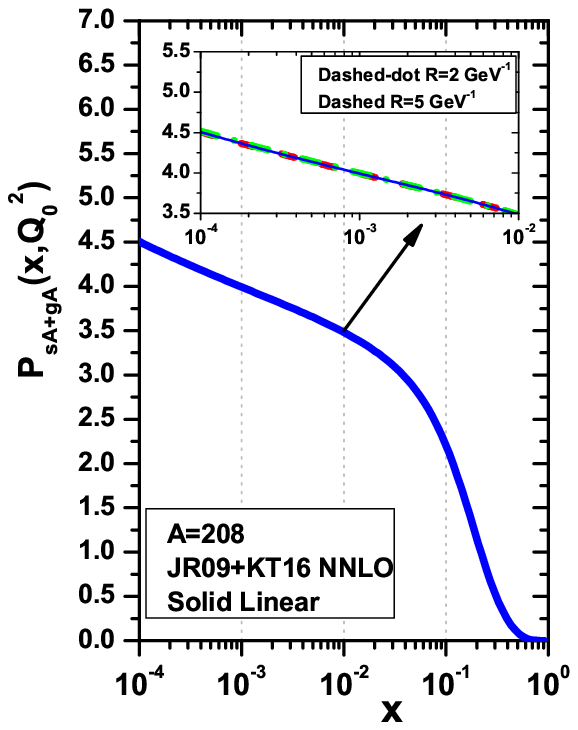}}
\caption{$P_{sA+gA}(x,Q_{0}^{2})$ as a function of $x$ on the
initial scale $Q_{0}^{2}=2~\mathrm{GeV}^2$ for the JR09 [30] and
KT16 [21] parameterization models in the NNLO approximation of the
nucleus of Pb-208(A=208,Z=82). The inset: the nonlinear
corrections compared with the linear (solid curve) at
$R=5~\mathrm{GeV}^{-1}$ (dashed curve) and $R=2~\mathrm{GeV}^{-1}$
(dashed-dot curve) for $x{\leq}10^{-2}$.}\label{Fig4}
\end{figure}
\begin{figure}
\centerline{
\includegraphics[width=0.5\textwidth]{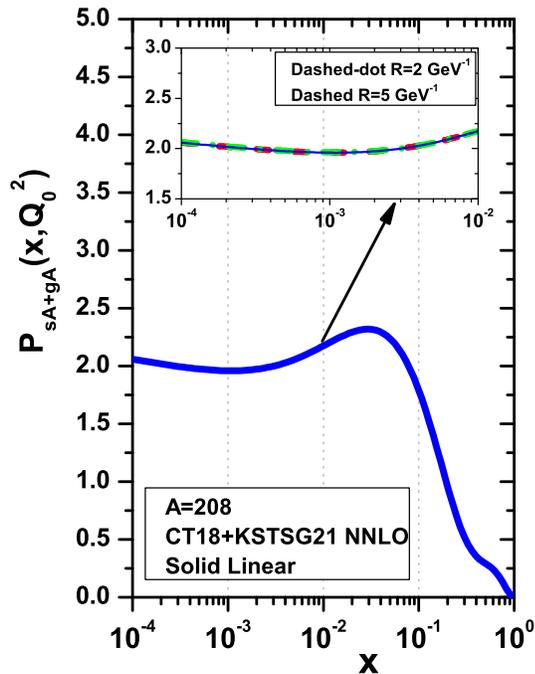}}
\caption{ $P_{sA+gA}(x,Q_{0}^{2})$ as a function of $x$ on the
initial scale $Q_{0}^{2}=1.69~\mathrm{GeV}^2$ for the CT18 [28]
and KSTSG21 [29] parameterization models in the NNLO approximation
of the nucleus of Pb-208(A=208,Z=82). The inset: the nonlinear
corrections compared with the linear (solid curve) at
$R=5~\mathrm{GeV}^{-1}$ (dashed curve) and $R=2~\mathrm{GeV}^{-1}$
(dashed-dot curve) for $x{\leq}10^{-2}$.}\label{Fig5}
\end{figure}
The nonlinear
($\xi_{A}^{{\mathrm{NLC}}}(x,x_{0},Q_{0}^{2}){\neq}1$) corrections
are compared with the linear in the region $x{\leq}10^{-2}$ for
both $R=2~\mathrm{GeV}^{-1}$(dashed-dot curves) and
$R=5~\mathrm{GeV}^{-1}$(dashed curves) for the light and heavy
nuclei in Figs.3-5. In these figures (i.e., Figs.3-5) we observe
that the violation of the nonlinear kernel
$P^{\mathrm{NLC}}_{sA+gA}(x,Q_{0}^{2})$ from the linear behavior
is observable for the light nuclei at low $x$ at the hotspot point
$R=2~\mathrm{GeV}^{-1}$. These violations of the linear behavior
are visible at low $x$ for light nuclei, although they are small
(see Fig.3), and they are invisible for heavy nuclei independent
of the
parametrization groups (see Figs.4 and 5).\\
\begin{table}[h]
\centering \caption{$I_{sA+gA}$ on the initial scale
$Q_{0}^{2}=2~\mathrm{GeV}^2$ for the JR09 and KT16 [21]
parameterization models [30] and on the initial scale
$Q_{0}^{2}=1.69~\mathrm{GeV}^2$ for the CT18 and KSTSG21 [29]
parameterization models [28]  in the NNLO approximation for the
nuclei of  C-12(A=12,Z=6) and Pb-208(A=208,Z=82) respectively. The
nonlinear corrections obtained at $R=5~\mathrm{GeV}^{-1}$ and
$R=2~\mathrm{GeV}^{-1}$ respectively. The differences
($\Delta^{A}=I^{\mathrm{Linear}}_{sA+gA}-I^{\mathrm{Nonlinear}}_{sA+gA}(R)$)
between the linear and nonlinear corrections for the singlet and
gluon contributions in the momentum sum rule for the light and
heavy nuclei are determined. }\label{table:table1}
\begin{minipage}{\linewidth}
\renewcommand{\thefootnote}{\thempfootnote}
\centering
\begin{tabular}{|l||c|c|c|} \hline\noalign{\smallskip} $I_{sA+gA}$ & JR09 & CT18 & CT18   \\
\hline\noalign{\smallskip}
Nuclei                              & A=208 & A=12 & A=208  \\
\hline\noalign{\smallskip}
Linear                              & 0.5939293 & 0.5562324 & 0.5619174  \\
Nonlinear($R=5~\mathrm{GeV}^{-1}$)          & 0.5939323 & 0.5562545 & 0.5619200  \\
Nonlinear($R=2~\mathrm{GeV}^{-1}$)          & 0.5939482 & 0.5563717 & 0.5619338   \\
$\Delta^{A}$($R=5~\mathrm{GeV}^{-1}$)          & -0.30${\times}10^{-5}$ & -0.0000221 & -0.26${\times}10^{-5}$   \\
$\Delta^{A}$($R=2~\mathrm{GeV}^{-1}$)          & -0.0000189 & -0.0001394 & -0.0000165   \\
\hline\noalign{\smallskip}
\end{tabular}
\end{minipage}
\end{table}
The deviations of the MSR, according to the nonlinear corrections,
for light and heavy nuclei are shown in Table III. In Table III,
we compared the linear and nonlinear corrections to the
$I_{sA+gA}$ for the heavy nucleus of Pb-208 on the initial scale
$Q_{0}^{2}=2~\mathrm{GeV}^2$ where the nPDFs are obtained from the
JR09 [30] set of the free proton PDFs by  weight functions
$w_{i},i=s,g$ are obtained from the KT16 [21] set of the nPDFs.
Also, the linear and nonlinear corrections to the $I_{sA+gA}$ for
the heavy nucleus of Pb-208 and the light nucleus of C-12 on the
initial scale $Q_{0}^{2}=1.69~\mathrm{GeV}^2$ are illustrated in
Table III, where the nPDFs are obtained from the CT18 [28] set of
the free proton PDFs by weight functions $w_{i},i=s,g$ are defined
from the KSTSG21 [29] set of the nPDFs. The nonlinear corrections
and their differences are obtained at
$R^{A}=A^{1/3}{\times}5~\mathrm{GeV}^{-1}$ and
$R^{A}=A^{1/3}{\times}2~\mathrm{GeV}^{-1}$ respectively. These
differences in $I_{sA+gA}$ can be considered in the light nuclei
at the hotspot point, but they are very small in the heavy nuclei.
Additionally, we furnish several predictions  of the ratio
$R=\frac{I_{sA+gA}}{AI_{s+g}}$ in the JR09 and CT18 sets  for the
nuclei C-12 and Pb-208 in Table IV, to be probed at upcoming
collider experiments such as EIC, as they are expected to improve
the precision of MSR at low-$x$.\\
\begin{table}[h]
\centering \caption{Ratio $R=\frac{I_{sA+gA}}{AI_{s+g}}$ for the
JR09 and CT18 sets in the NNLO approximation for the nuclei of
C-12 and Pb-208 at $R=5~\mathrm{GeV}^{-1}$ and
$R=2~\mathrm{GeV}^{-1}$. }\label{table:table1}
\begin{minipage}{\linewidth}
\renewcommand{\thefootnote}{\thempfootnote}
\centering
\begin{tabular}{|l||c|c|c|} \hline\noalign{\smallskip} $R=\frac{I_{sA+gA}}{AI_{s+g}}{\times}10^{-2}$ & JR09 & CT18 & CT18   \\
\hline\noalign{\smallskip}
Nuclei                              & A=208 & A=12 & A=208  \\
\hline\noalign{\smallskip}
Linear                              & 0.52380 & 8.57640 & 0.49985  \\
Nonlinear($R=5~\mathrm{GeV}^{-1}$)          & 0.52370 & 8.57430 & 0.49971  \\
Nonlinear($R=2~\mathrm{GeV}^{-1}$)          & 0.52317 & 8.57860 & 0.49987   \\
\hline\noalign{\smallskip}
\end{tabular}
\end{minipage}
\end{table}
In summary, we have investigated the effects of nonlinear
corrections to the momentum sum rule at the initial scale
$Q_{0}^2$. Using the shadowing effects at low $x$ and the known
parton distribution functions, we are able to add nonlinear
corrections to the momentum sum rule for the nucleons and nuclei.
Interestingly, these effects increase the description of the
singlet and gluon contributions to the momentum sum rule  at the
beginning of low $Q^{2}$ evolution. The main effect is to increase
the singlet and gluon contributions to the  momentum sum rule  for
the proton and the light nuclei at very low $x$ at the hotspot
point, which can be helpful for the
upcoming LHeC and EICs.\\
%%%%%%%%%%%%%%%%%%%%%%%%%%%%%%%%%%%%%%%%%%%%%%%%%%%%%%%%%%%%%%%%%%%%%%%%%%%%%%%%%%%%%%%%%%%%%%%
%%%%%%%%%%%%%%%%%%%%%%%%%%%%%%%%%%%%%

%%%%%%%%%%%%%%%%%%%%%%%%%%%%%%%%%%%%%%%%%%%%%%%%%%%%%%

\subsection{ACKNOWLEDGMENTS}

I am grateful to Razi University for the financial support of this
project.\\
% I am especially grateful to Stanley J. Brodsky for
%valuable discussion and critical remarks.\\
%%%%%%%%%%%%%%%%%%%%%%%%%%%%%%%%%%%%%%%%%%%%%%%%%%%%%%%%%%%%%%%%%%%%%%%%
%%%%%%%%%%%%%%%%%%%%%%%%%%%%%%%%%%%%%%%%%%%%%%%%%%%%%

%%%%%%%%%%%%%%%%%%%%%%%%%%%%%%%%%%%%%%%%%%%%%%%%%%%
\section{References}
1. M.A.Shifmann, A.I.Vainshtein and V.I.Zakharov, Nucl.Phys.B
{\bf147}, 385 (1979).\\
2. W.Greiner and A.Schafer, Quantum Chromodynamics, Springer
1994.\\
3. K.Golec-Biernat and Anna M.Stasto, Phys.Rev.D{\bf107},
054020 (2023); S.J.Brodsky, I.Schmidt and S.Liuti, arXiv[hep-ph]:1908.06317.\\
4. G.Dissertori, I.Knowles and M.Schmelling, Quantum
Chromodynamics, High Energy Experiments and Theory, Oxford
University Press (2003).\\
5. R.G.Roberts, The Structure of the proton, Deep Inelastic
Scattering, Cambridge University Press (1990).\\
6. S.L.Adler, Phys.Rev. {\bf143}, 1144 (1966).\\
7. J.D.Bjorken, Phys.Rev. {\bf163}, 1767 (1967).\\
8. D.J.Gross and C.H. Llewellyn Smith, Nucl.Phys.B{\bf14}, 337
(1969).\\
9. K.Gottfried, Phys.Rev.Lett.{\bf18}, 1154 (1967).\\
10. M.Botje, Eur.Phys.J.C {\bf14}, 285 (2000).\\
11. L. V. Gribov, E. M. Levin and M. G. Ryskin, Phys. Rept.
{\bf100}, 1 (1983).\\
12. A. H. Mueller and J. w. Qiu, Nucl. Phys. B {\bf268}, 427
(1986).\\
13. M.R.Pelicer et al., Eur.Phys.J.C {\bf79}, 9 (2019).\\
14. V.A.Abramovsky, V.N.Gribov, O.V.Kancheli, Yad. Fiz. {\bf18},
595 (1973); Sov.J.Nucl.Phys. {\bf18}, 308 (1974).\\
15. J.Kwiecinski et al., Phys.Rev.D {\bf42}, 3645 (1990).\\
16. J.Collins and J.Kwiecinski, Nucl.Phys.B {\bf335}, 89 (1990).\\
17. M.Lalung, P.Phukan and J.K.Sarma, Nucl.Phys.A {\bf984}, 29
(2019); G.R.Boroun, arXiv:2305.04243; G.R.Boroun and B.Rezaei, Eur.Phys.J.C {\bf81}, 851 (2021).\\
18. I.Balitsky, Nucl.Phys.B {\bf463}, 99 (1996).\\
19. Y.V.Kovchegov, Phys.Rev.D {\bf60}, 034008 (1999).\\
20. M.Hirai, S.Kumano and M.Miyama, Phys. Rev. D {\bf64}, 034003 (2001).\\
21. H.Khanpour and S.Atashbar Tehrani, Phys.Rev.D {\bf93}, 014026
(2016).\\
22. H.Khanpour et al., Phys.Rev.D {\bf 104}, 034010 (2021);
S.Atashbar Tehrani, Phys.Rev.C {\bf86}, 064301 (2012).\\
23. M. Hirai, S. Kumano and T.-H. Nagai, Phys. Rev. C {\bf70},
044905 (2004).\\
24. J.Rausch, V.Guzey and M.Klasen, Phys.Rev.D 107, 054003 (2023);
G.R.Boroun, B.Rezaie and F.Abdi, arXiv:2305.01893; G.R.Boroun and
B.Rezaei, arXiv:2303.07654; Phys.Rev.C {\bf107}, 025209 (2023); F. Muhammadi and B.Rezaei, Phys.Rev.C {\bf106},  025203(2022).\\
25. LHeC Collaboration, FCC-he Study Group, P. Agostini, et al.,
J. Phys. G, Nucl. Part. Phys. {\bf48},  110501 (2021).\\
26. R.Abdul Khalek et al., Nucl. Phys.A {\bf 1026}, 122447
(2022).\\
27. A.D.Martin, W.J.Stirling, R.S.Thorne and G.Watt,
Eur.Phys.J.C {\bf63}, 189 (2009).\\
28. Tie-Jiun Hou et al., Phys.Rev.D {\bf103}, 014013 (2021).\\
29. H.Khanpour et al., Phys.Rev.D {\bf104}, 034010 (2021).\\
30. P.Jimenez-Delgado and E.Reya, Phys.Rev.D {\bf79} , 074023
(2009).\\
31. G.C.Nayak, arXiv [hep-ph]:1804.02712; 1806.01220.\\
32. S.J.Brodsky, V.E.Lyubovitskij and  I.Schmidt, Phys.Lett.B
{\bf824}, 136812 (2022).\\

%%%%%%%%%%%%%%%%%%%%%%%%%%%%%%%%%%%%%%%%%%%%%%%%

%\begin{figure}
%\includegraphics[width=1\textwidth]{Fig1}
%\caption{The linear and non-linear gluon distribution function at
%$R=2~\mathrm{GeV}^{-1}$ for $Q^{2}=10, 30, 50$ and
%$100~\mathrm{GeV^{2}}$ with respect to the parametrization model
%[7] and GLR-MQ equation [36] respectively.}\label{Fig1}
%\end{figure}

%%%%%%%%%%%%%%%%%%%%%%%%%%%%%%%%%%%%%%%%%%%%%%%%%

%%%%%%%%%%%%%%%%%%%%%%%%%%%%%%%%%%%%%%%%%%%%%%%%%%%%%%%
%%%%%%%%%%%%%%%%%%%%%%%%%%%%%%%%%%%%%%%%%%%%%%%%%%%%%%%%
\end{document}